\documentclass[draft]{aipproc}
\layoutstyle{6x9}
\usepackage{times}
\usepackage{amsmath}
\usepackage{amsfonts}
\usepackage{amssymb}
\usepackage{graphicx}
\usepackage{afterpage}

\begin{document}

\title{Critical Thoughts on Cosmology}

\classification{    }

\keywords      {Hawking entropy, Hawking radiation, cold Big Bang, dark
energy, dark matter, supernovae, cosmic rays, $\gamma$-ray bursts, jets, burning disks, anomalous redshifts.}

\author{Wolfgang Kundt}{address={Argelander Institute for Astronomy of Bonn University, Auf dem H\"{u}gel 71, D-53121 Bonn, Germany }}

\begin{abstract}
An overview is given in section 1, of uncertain building blocks of present-day
cosmologies. Thereafter, these edited lecture notes deal with the following
four special problems: (1) They advertise Wiltshire's result --\ making `dark
energy' obsolete -- that accelerated cosmic expansion may be an artefact, due
to an incorrect evaluation of the cosmic timescale in a Universe whose bulk
matter is inhomogeneously distributed. (2) They cast doubt on Hawking's
prediction of black-hole evaporation. (3) They point at various
inconsistencies of the black-hole paradigm, in favour of nuclear-burning
central engines of AGN. (4)\ They re-interpret (a best case of)\ `anomalous
redshifts' as non-cosmological, kinematic redshifts in strong jet sources.
\end{abstract}

\maketitle

\section{1. BASICS\ OF\ COSMOLOGY}

The literature on Cosmology is nowadays quite heterogeneous; how certain are we
concerning its basic assumptions? When we try to explore our cosmic past by
evaluating all the astronomical observations, our confidence is strengthened
by the fact that:

(0) All the dimension-less \emph{fundamental constants} have been constant
throughout cosmic epochs (as judged by their redshift of recession), at a
level of $\lesssim10^{-5}$ (Kanekar et al, 2005). I.e. we feel encouraged to
apply to cosmology our locally secured laws of physics. We then have to worry
about the proper field equations:

(1) Should cosmology be based on Einstein's \emph{Theory of General
Relativity}, with or without the cosmological ($\Lambda$) term, called ''dark
energy'' in more modern language? Authors like David Crawford (2008), Wilfred Sorrell (2006), or Tom
van Flandern prefer \emph{Newtonian} cosmologies. Hoyle et al (2000) think
they require \emph{continuous creation} of matter, at near-singular sites.
Does the Universe contain ''\emph{dark} (non-baryonic) \emph{matter}'', as is
generally believed \ --  not necessarily, though, by Erwin de Blok (McGaugh
and de Blok, 1998)? Authors like David Wiltshire (2007a,b) argue that the
mystery of \emph{dark energy} is not required once we evaluate our backward
lightcones correctly, taking care of the (observed) inhomogeneous distribution
of the field-generating matter; see also Ellis (2008). 

(2) Once we agree on the field equations -- with or without a certain number
of free parameters -- there is the unknown initial state: Was there a ''big
bang'' singular beginning? Was it \emph{hot} -- as is usually assumed, for no
other than simplicity's sake -- or was it \emph{cold}, as preferred by David
Layzer (1990)? To me, a cold beginning sounds like the \emph{most plausible}
initial condition. Moreover, I expect a large fraction of the `primordial'
cosmic helium to be formed in the central engines of all the galaxies during
their active epochs, arguing against excessive primordial nuclear burning of
hydrogen (during the first three minutes; Kundt 2008a).

(3) Within this cosmological framework of assumptions, we want to understand
how the network of cosmic \emph{density fluctuations} was formed -- of `voids'
surrounded by `walls' -- with its \emph{galaxies}, and \emph{clusters} of
galaxies within which large-scale \emph{magnetic fields} and \emph{stars}
form, shine, blow winds, and explode, and whose centers turn
(non-thermally)\ active quasi-periodically, with bursts of star formation, and
with\ \emph{QSOs} at their very centers, with their Broad-Line Regions (BLRs),
and occasional gigantic twin jets, Narrow-Line Regions (NLRs), and Extended
Shell Regions (ESRs). How far has our knowledge about them advanced; are we
ready to cope with them? The building blocks contain (the mechanism of the
various) \emph{supernova explosions} (Kundt 2008b), \emph{jet formation}
(Kundt \& Krishna 2004), \emph{cosmic-ray} production (by the galactic
throttled pulsars? Kundt 2009), \emph{gamma-ray bursts} (again by the
throttled pulsars? Kundt 2009), anomalous (\emph{non-cosmological?})
\emph{redshifts} (Arp, 2008), and a thorough understanding of the
\emph{fluctuation structure} of the (2.725 K) background radiation: How much
of it is imprinted by foreground structure, most noticeably by the solar
system, as is suggested by its quadrupole and octupole moment (Thyrso Villela,
these proceedings; but also Fixsen 2003)? See also Kundt (2005) for preferred interpretations.

(4) Literature on cosmology does not only deal with the items listed in
paragraph (3), but also with the possible formation of \emph{black holes},
with their \emph{entropy}, and with their \emph{radiation} (in particular when
of low mass), (Carroll 2008). None of them may be realistic, or even rightly
claimed (Belinski 2006, Leblanc 2002); I will come back to them in section 3.
Carroll also talks about a \emph{quantization of space-time geometry}; with
what (measurable) effects in mind? I cannot see any astronomical observation
that would (be able to) measure space-time quantization. Astronomical
observations never approach the frontier between classical and quantum
behaviour of its targets. Independently, I cannot see a consistent fusion of
the two theories, cf. (Kundt 2007). Quantization celebrates its successes
whenever the sizes of particles shrink inside the ranges of their guiding
waves. It controls the equation of state of the substratum, but should leave
the spacetime metric unquantized.

This compilation of edited lecture notes will focus on a number of
contoroversial items which are of relevance to modern cosmology, as has just
been explained. \emph{Section 2} will present an intuitive explanation of how
Wiltshire has rendered\ `\emph{dark energy}' obsolete: smoothening the
spacetime geometry does not commute with evaluating its past lightcones, and
timing. \emph{Section 3} repeats my earlier (1976) objection to Hawking's
definition of the term `\emph{BH entropy}': his expression measures the
entropy of the BH's evaporation products (if it dissolved via radiation), not
that of a newly formed BH. \emph{Section 4} summarizes my lack of conviction
of the presence of supermassive black holes at the centers of all the (large)
galaxies, in favour of (nuclear)\ `\emph{burning disks}' (BDs). And
\emph{section 5} deals with the phenomenon of \emph{anomalous redshifts} which
have stood at the cradle of non-Big-Bang cosmologies.

\section{2. TIME\ KEEPING\ IN\ AN\ INHOMOGENEOUS\ UNIVERSE}

`Dark energy' is the name introduced by Mike Turner, for what had been called
the `$\Lambda$ term', or\ `cosmological term' in Einstein's field equations
for more than half a century, a term that had no obvious physical meaning --
at least not in the laboratory -- but that could not be rejected either from
the cosmological field equations if one was looking for the most general
second-order equations derivable from a scalar Lagrangean. During the last
decade, measurements with increasing accuracy of the present average cosmic
expansion signalled an increasing expansion rate of the substratum -- an
acceleration -- in obvious violation of energy conservation: An expanding
cloud of self-gravitating objects should decelerate. This misbehaviour of
cosmological kinematics urged Turner to introduce his cryptic -- and even
somewhat misleading -- name ''dark energy'' for the $\Lambda$-term: $\Lambda$
does not correspond to an energy density because it exerts a negative
pressure, forbidden by the classical energy inequalities for laboratory
substance (e.g. Kundt 1972); it is a non-energy, or at best a quasi-energy.

For this reason, it struck me as a salvation of (serious) cosmology when I
read about David Wiltshire's dismissing dark energy (Wiltshire 2007a,b, Ellis
2008). His thesis is simple and convincing: Cosmology had hitherto been
evaluated wrongly, by ignoring the inhomogeneous distribution of its
substratum. We know Shapiro's `time delay' effect in the solar system, and in
close neutron-star binaries: Signals passing close to heavy objects (stars,
galaxies) reach a distant observer with a certain delay. In the same vein,
when we measure cosmic expansion, we use light rays which have propagated
through an inhomogenous Universe, with voids and walls, sometimes propagating
through near-vacuum patches (voids), and sometimes skimming heavy mass
concentrations (clusters of galaxies, in the walls). Clearly, the formulae
derived for a homogeneous cosmological model cannot be expected to describe
our observations correctly, due to non-linearities. Our local time scale,
described by our (timelike) worldline, inside our (massive) Galaxy, has to be
referred to the average cosmic timescale via intersections with successive
null geodesics lying on past light cones, and connecting us to distant sources
in the past. There is no a priori reason why these two timescales should be
the same. A deviation is expected, an acceleration, whose sign we must
calculate, and whose magnitude must likewise be calculated. It is a cumulative
effect, to be obtained by integration over large spacetime distances.
Wiltshire has done such calculations, and claims that their result describes
the observed seemingly accelerated expansion, without a $\Lambda$-term in the
field equations. All we have to do is evaluate our observations rigorously.


\begin{figure}
\centering
\includegraphics[width=12.0cm]{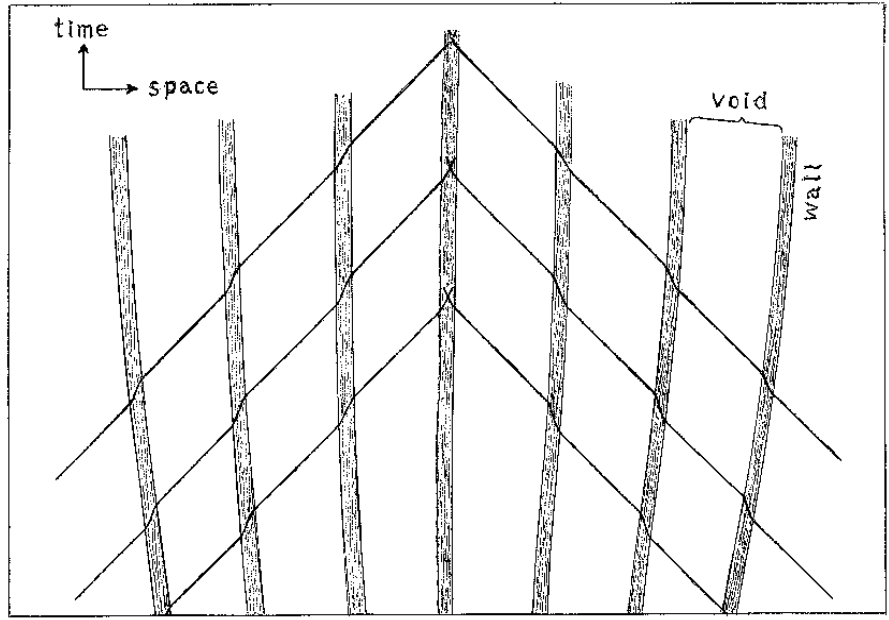}
\caption { (1+1)-dim radial spacetime section of our past cosmic lightcone, in
non-Minkowskian coordinates: Matter is concentrated in non-expanding
(timelike) worldtubes separated by expanding cosmic voids. The variable delay
of cosmic time inside world tubes is indicated by wiggles in the past global
light cones. }
\end{figure}


Wiltshire's papers are not easy to read; they are long. But fig.1 should do in
explaining what he has done: It sketches a significant fraction of our cosmic
environment, in an almost metrical (1+1)-dim spacetime slice through our
Universe, whose metric is indicated -- upto an arbitrary conformal distortion
-- by a number of past lightcones, with their tips at the center of our local
world tube (of higher than average mass density). These lightcones are steeper
when traversing the walls, and shallower in between, because signals propagate
more slowly -- as sensed by a distant observer -- when they move through more
densely populated domains than otherwise. Precisely this locally inhomogeneous
geometry gives rise to a non-trivial global effect, when measuring our
(average) past spacetime geometry. No dark mystery is required for its description.

%

\section{3. ENTROPY OF\ A\ BLACK\ HOLE}

Let me begin this section with a (2+1)-dim sketch of the spacetime geometry of
a forming (non-rotating) BH which is assumed to subsequently dissolve again by
heating up, radiating, shrinking, and finally exploding. The BH is assumed to
form from an approximately spherical (supercritical) mass concentration via
collapse under its own gravity. Similar in spirit to fig.1, fig.2 is drawn in
asymptotically (2+1)-dim Minkowskian coordinates, assuming spherical symmetry
of 3-space, but metrically distorted near its center in such a way that the
causal structure has to be read off the drawn-in local light cones, which
point increasingly inward during increasing approach of the symmetry axis of
the figure. This symmetry axis represents the history of the forming BH's
center, which at late times -- after the BH's assumed complete evaporation --
turns again into the center of a Minkowskian domain. During collapse, the
contracting substratum gives off all the higher multipole moments of its mass
distribution via radiation (of both electromagnetic and gravitational waves),
and contracts deeply inside its `horizon', which is drawn (in gray) in the
shape of a slowly contracting (lightlike) cylinder. A distant observer sees the
surface of the contracting mass concentration until it crosses its horizon.
Thereafter, he or she receives the shrinking hole's redshifted evaporation
radiation, for a very long time, whose mass decreases slowly -- towards
$\lesssim10^{9}$g -- and whose temperature rises slowly, and eventually peaks
abruptly, above $10^{17}$K, in the form of a final flash, of duration of the
order of a second.

\begin{figure}
\centering
\hspace{3mm}
  \begin{minipage}[b]{7.5cm}
  \includegraphics[width=70mm]{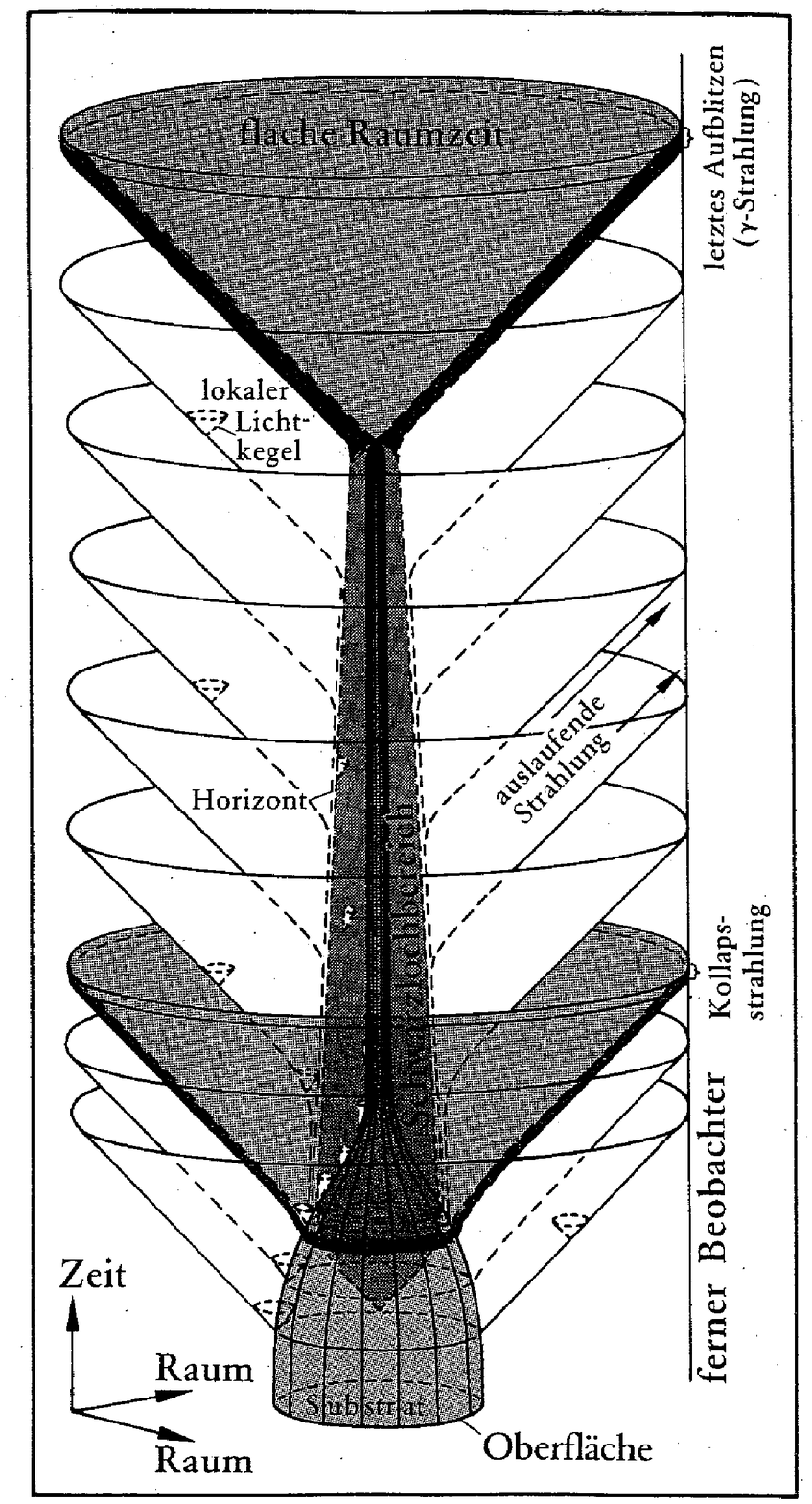}
   \parbox[b]{70mm}{\small \  }
   \parbox[b]{70mm}{\small \textbf{FIGURE 2.} (2+1)-dim spacetime diagram of the history of a forming, and subsequently evaporating stellar-mass BH, according to Hawking's prediction (1974, 1975). Spherical symmetry is assumed, and coordinates are chosen Minkowskian at large distances from the center, whilst strong distortions near the center are indicated by (small) local lightcones. In these coordinates, the BH domain proper is the dark-gray elongated central al\--
 }
  \end{minipage}%
  \begin{minipage}[b]{7.5cm}
  \includegraphics[width=70mm]{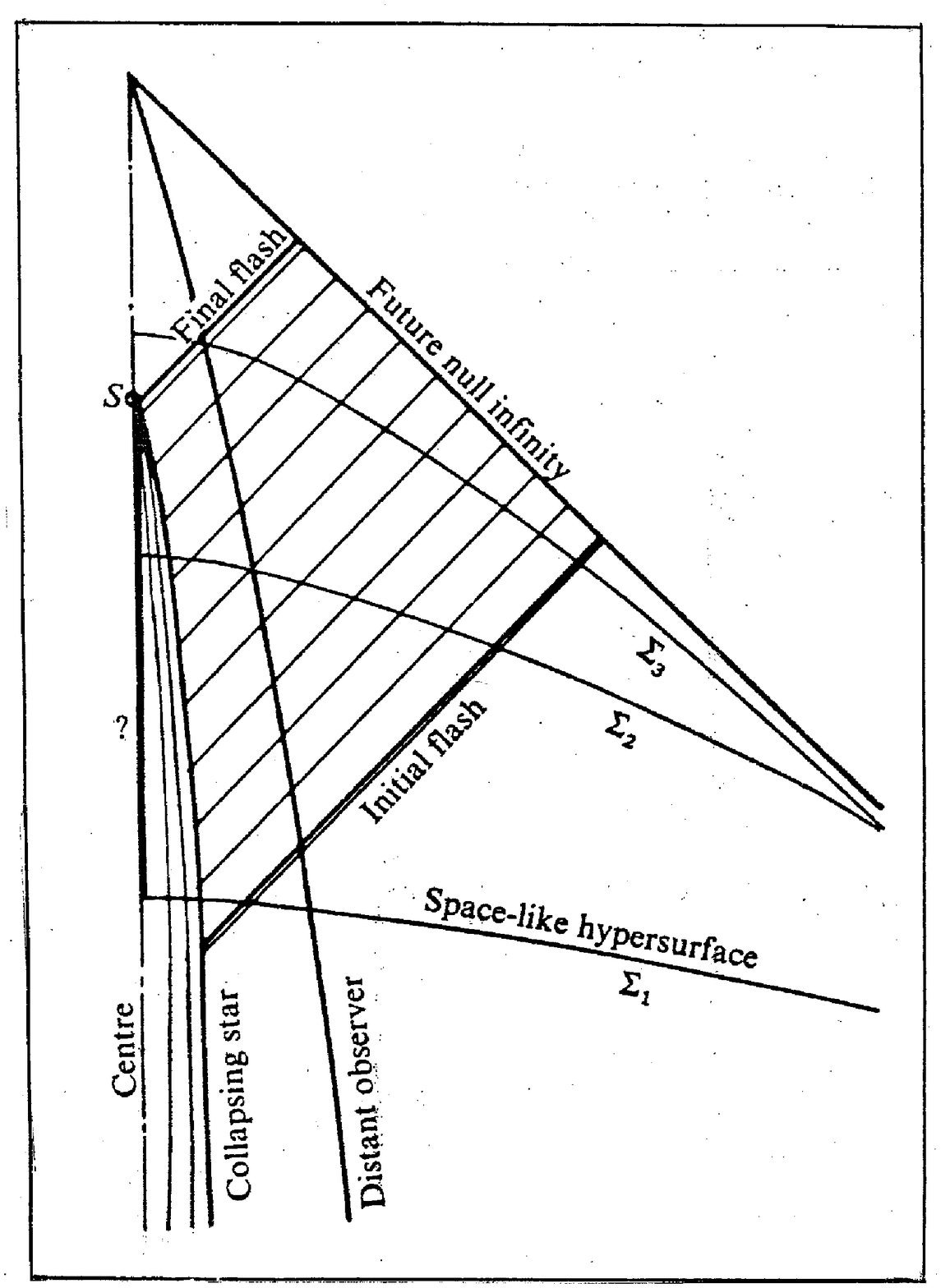}
   \parbox[b]{70mm}{\small \  }
   \parbox[b]{70mm}{\small
most cylinder, which
terminates during the final flash. A distant observer sees the BH formation
via its very short burst of radiation during formation, when all the
non-fitting higher multipole moments are disposed of, then via its extremely
faint evaporation radiation, for almost eternal times, and eventually via its
short, very hot final flash of disintegration. \newline  
\textbf{FIGURE 3.} (1+1)-dim radial spacetime section through the geometry of fig.2, now
in conformally distorted (Penrose) coordinates for which future null infinity
has been transformed to finite distances, and all lightrays propagate at
$\pm45^{0}$. A set of spacelike hypersurfaces $\Sigma_{j}$ is drawn, to which
a distant observer would refer his or her entropy estimates.}
  \end{minipage}%
\end{figure}

\afterpage{\clearpage}

This history of a BH just described, and sketched in figs.2,3, was advocated by
Stephen Hawking in 1974, and elaborated by him in 1975, and we all trusted it,
throughout the world. We trusted him and his associates, even though we did
not understand the -- highly non-classical -- mechanism by which some strongly
curved spacetime domain (around the BH) could generate outgoing
electromagnetic radiation, and cause its enclosed volume to shrink in mass,
size, and inverse temperature. Only now at this School, 34 years later, do I
learn from Belinski (2006) and Leblanc (2002, recited) that all these
expectations may have been premature, and unrealistic. That a BH, should it
form, will not have a temperature, and will not evaporate. It will just sit
and wait and grow by accretion from its surroundings.

Even though in 1975, I trusted Hawking's BH evaporation scenario, I disagreed
with him on the meaning of what he called ''BH entropy''. My objections
appeared finally in print, in 1976, with 11 distorting printing errors, and
were mostly ignored by the scientific community. In that publication, I
compared a forming stellar-mass BH with a forming white dwarf, or neutron
star, and showed that all those compact stellar remnants (under collapse) had
small entropies, smaller than the material from which they had formed, and
that Hawking's so-called ''BH\ entropy'' agreed with that of the hole's
expected randomized evaporation product (after some 10$^{67}$yr), a huge bath
of radio waves of wavelength some 20 Km. Ever since then, string theorists
have been proud of being able to rederive this expression, not worrying about
its physical meaning. As Constantin Tsallis has shown, there exist large
classes of functions with the (reasonable) positivity and convexity properties
of the standard entropy in thermodynamics, though violating additivity (Boon
and Tsallis, 2005; they use the word ''nonextensive'' for ''non-additive'').
Hawking's is one of them; it is quadratic in the BH's mass, not linear. The
relevant thermodynamic formulae will soon follow.

\label{sec:dis}Before their presentation, it will be helpful to redraw fig.2
\ in a different (distorted) way, leaving the local lightcones at $\pm45^{0}$.
Such conformally distorted diagrams can map infinity onto finite surfaces;
they leave spacelike surfaces weakly inclined ( $< 45^{0}$), and timelike 
surfaces strongly inclined ( $> 45^{0}$) w.r.t. the time axis. Fig.3 is a redrawing of fig.2, but only for
(1+1)-dim meridional sections. It shows the set of spacelike hypersurfaces
$\Sigma_{j}$ for whose material contents I shall calculate the successive
entropies $S_{j}$ contained in them. Note that quantum cosmology proposes yet
different expressions -- likewise called ''entropy'' -- which do not vanish
for vanishing particle number densities (Carroll 2008); I do not understand
their physical meaning. They violate the strong equivalence principle.

We are now ready to calculate the relevant entropies $S_{j}$. Independently of
whether we choose the phenomenological approach of box thermodynamics, with
$dS$ := ($dU$ + $pdV$)/$kT$, with $U$ standing for internal energy, $p$ :=
pressure, and $V$ := volume, or the statistical mechanics approach \ $S/
Nk$ := $-<W,ln(W)>$ , with $N$ := number of particles, and $W$ := the
canonical equilibrium distribution for a homogeneous gas of number density $n
$ at temperature $T$, the textbooks tell us that%

\begin{equation}
S=N\text{ }k\text{ }s\text{ \ \ \ with \ \ \ \ }s=5/2+\ln(1\text{ }/\text{
}n\text{ }\lambda_{th}^{3})\text{ \ \ \ }\in\text{ \ \ ( 0 , 90
)}\label{entropy}%
\end{equation}
holds for a non-quantum, non-relativistic (hydrogen) gas whose thermal de
Broglie wavelength\ reads$\ \lambda_{th}$ : = $h$ $/\sqrt{2\pi mkT}$ =
$10^{-10.3}$cm $/\sqrt{T_{7}}$ \ with $m$ = $m$(proton). Note that for
ordinary matter, the entropy density $s$ takes small values, between 1 and 90
for non-quantum gases, but always positive, and never very large values; it
can be considered of order unity in astrophysical applications. \ This formula
can be easily generalized to Newtonian gases in a curved spacetime by
integrating the entropy density $s$ , moving with 4-velocity $u^{a}$, over a
space section $\Sigma$ of differential 3-volume \ $d\overset{\ast}{x_{a}}$ :
\begin{equation}
S(\Sigma)\text{ \ }=\text{ }\underset{\Sigma}{\int}s^{a}\text{ \ }d\overset
{}{\overset{\ast}{x_{a}}\text{ \ \ }}\text{where \ \ \ }s^{a}\text{ :=
}s\text{ }u^{a}\text{ .}\label{integral}%
\end{equation}
For a hydrogen mass $M$ inside $\Sigma$, (\ref{integral}) yields \ $S(M)$ =
$10^{57}$ $k$ $(M\ /$ $M_{\odot})$ $s(M)$ $\ \lesssim$ $10^{59}$ $k$ $(M$ $/$
$M_{\odot})$ .

These expressions are to be compared with Hawking's entropy expression for a
non-rotating black hole of mass $M$, (Schwarzschild) radius $R$ = $2GM/
c^{2}$ = 10$^{5.5}$cm ($M/M_{\odot}$) , and temperature
\begin{equation}
T_{BH}:=\hbar\text{ }c^{3}\text{ }/\text{ }8\pi GM\text{ }k\text{ }%
=10^{-7}\text{K }(M_{\odot}\text{ }/\text{ }M)\text{ ,}\label{temperature}%
\end{equation}
which imply an evaporation time $t_{ev}$ for blackbody radiation given by
\begin{equation}
t_{ev}=Mc^{2}\text{ }/\text{ }4\pi R^{2}\text{ }\sigma_{SB}\text{ }T_{BH}%
^{4}\approx10^{67}\text{yr }(M\text{ }/\text{ }M_{\odot})^{3}\text{
.}\label{evaporation}%
\end{equation}
This evaporation time shrinks to $t_{ev}$ =$1$sec for $M$ shrinking to
$10^{8.5}$g, at a BH temperature of $T_{BH}$ = $10^{17.8}$K , higher than any
(effective) temperature reached yet in laboratory experiments, and therefore
to be handled with some reservation. Still, order-of-magnitude-wise, it
describes the general expectations since the late 70s. Hawking's entropy
expression for a BH reads:
\begin{equation}
S_{BH}=4\pi GM^{2}\text{ }k\text{ }/\text{ }\hbar\text{ \ }\approx\text{
}10^{77}k\text{ }(M\text{ }/\text{ }M_{\odot})^{2}\text{ .}\label{Hawking}%
\end{equation}
When divided by above entropy (\ref{integral})\ of its constituent hydrogen
mass, with $s(M)$ $\lesssim10^{2}$, it yields the announced result:
\begin{equation}
S_{BH}\text{ }/\text{ }S(M)=(m_{p}c^{2}/\text{ }h\nu_{\odot})(2\pi\text{
}/\text{ }s(M))(M\text{ }/\text{ }M_{\odot})\approx10^{19}\text{ }(M\text{
}/\text{ }M_{\odot})\text{ ,}\label{ratio}%
\end{equation}
in which $\nu_{\odot}$ := $c/2\pi R_{\odot}$ = $10^{4.2}$Hz stands for
the peak frequency of a solar-mass BH's decay radiation. The huge factor
$(m_{p}c^{2}/$ $h\nu_{\odot})$ = $10^{19}$ (for a solar mass $M=M_{\odot}$)
measures the number of decay photons generated during the hole's 10$^{67}$
years of decay: $10^{19}$ radio photons for (the energy of) one
hydrogen atom. Clearly, this huge number has no physical relevance for a newly
formed BH, only for its eventual decay product.

As already stated above, this eventual decay product may never form (Leblanc
2002, Belinski 2006), because BHs do not evaporate. But in the meantime --
before this minority opinion has succeeded in replacing the textbook interpretation --
above quantitative results can serve as a warning: that untested QFT results
need not apply. Frontline physics need not always be reliable.

\section{4. CENTRAL ENGINES OF\ ACTIVE\ GALAXIES}

The brightest sources in the Universe are the central engines of (massive)
galaxies \ -- \ even with the GRBs included, which I purposely ignore in this
communication, (cf. Kundt 2009) -- \ whose luminosities can exceed those of
their host galaxies by factors of $\lesssim10^{2}$ . They are commonly thought
to be powered by supermassive black holes, of masses $\lesssim10^{10}%
$M$_{\odot}$, originally because of their huge radiative outputs, occasionally
dwarfing their hosts, already at optical frequencies, but even more so at TeV
photon energies. 

$\bullet$ But are we permitted to assume that supermassive black holes have
gigantic radiative efficiencies, of order $\lesssim$0.4, rather than
$\ll 10^{-3}$, expected on alternative grounds (see below)? $\bullet$ And how did those
black holes form in the first place?\ Have not centrifugal and pressure forces
always exceeded the self attraction of central galactic-mass accumulations?
$\bullet$ Why have the masses of the observed central BHs decreased during
cosmic epochs, from initial $\lesssim10^{9.5}$ M$_{\odot}$ to their
present-day $\lesssim10^{7}$M$_{\odot}$, as shown by the statistics of the
SDSSurvey (fig.4, Vestergaard et al 2008)? $\bullet$ How could some of the most
massive ones already form within $\lesssim0.8$ Gyr after the Big Bang? $\bullet$
Why do their masses scale as $10^{-2.85}$ times their bulge masses (Marconi
and Hunt 2003)? $\bullet$\ How do they blow their gigantic winds, and why have
those winds the chemistry of ashes from excessive nuclear burning, being
$\gtrsim10^{2}$-fold metal enriched (upto Fe)? $\bullet$\ How do they generate
their extremely hard spectra, (occasionally) peaking at $\gtrsim$TeV energies,
even recorded (from PKS 2155-304) as minute-sharp, hour-scale bursts (Weekes
2007), whilst accreting black holes radiating at their Eddington rates are
predicted to shine with blackbody temperatures of KeV($M_{\odot}/M$)$^{1/4}$?
$\bullet$\ Why are some of them distinctly underluminous? $\bullet$ Why does their
high $\gamma$-ray compactness not prevent them from forming jets, in the
($10\%$) cases of their radio-loud subpopulation, via inverse-Compton losses?
$\bullet$\ And, if all the astrophysical jet sources are generated by a
universal type of engine \ -- whose powerhouses are newly forming stars (like
our Sun, in its past), forming (binary) white dwarfs, binary neutron stars,
and AGN -- this universal type of engine looks like a rotating magnet, not
like a BH (Kundt and Krishna 2004).

\begin{figure}
\centering 
\begin{minipage}[b]{16.cm}
\centering
\includegraphics[width=16.0cm]{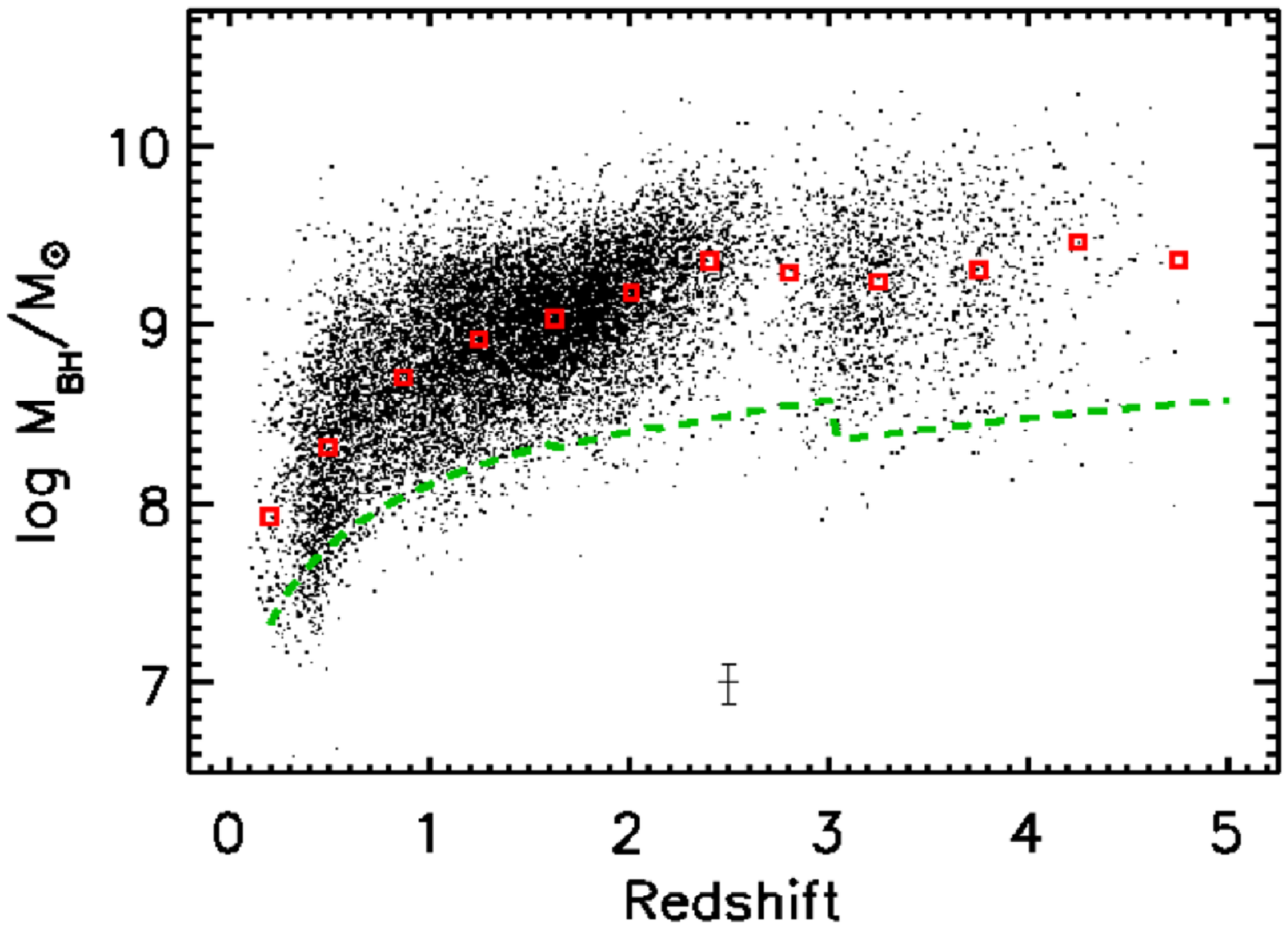} 
\parbox[b]{145mm}{\small \  }
\parbox[b]{145mm}{\small \textbf{FIGURE 4.} (Estimated) mass distribution of 14,584 quasar central engines (CEs)
with z $\geq$ 0.2, as functions of redshift z, from the Sloan Digital Sky
Survey Data Release 3, within an effective sky area of 1644 deg$^{2}$, taken
from Vestergaard et al (2008). Squares denote median masses in each redshift
bin. The dashed curve indicates faint SDSS flux limits. }
\end{minipage}
\end{figure}

None of these questions have ever been satisfactorily answered in the
literature, as far as I know (Kundt 2002, 2008a). There always was the
seemingly unsolved problem of the required energetics, thought to exceed the
nuclear reservoir provided by the primordial hydrogen. This problem is absent
in David Layzer's cold Big Bang approach. Explosive nuclear burning can take
care of the gigantic mass ejections from the centers of galaxies, evidenced in
the form of the BLR, NLR, and ESR, so that the CEs of the QSOs started massive
at high redshifts, at $\lesssim10^{10}$M$_{\odot}$, were repeatedly discharged
during active cycles (of their hosts), and have presently shrunken to their
(statistically) low masses of $\lesssim10^{7}$M$_{\odot}$, (fig.4). In this
process, their metallicities will have grown steadily, via incomplete ejection
of ashes, so that present-day activities occur at distinctly lower masses of
the CEs; which I like to call BDs, ''burning disks'', or ''flat stars''. I
conceive them as the continuous continuations of the well-known gaseous galactic
disks, all the way to their centers, cf. fig.5. During spiral-in \ -- at mass
rates of $\lesssim$ M$_{\odot}$/yr, roughly radius-independent for (large)
galactic disks \ -- matter accumulates in their centers until it reaches
stellar densities, starts main-sequence burning, and eventually heats up to
explosive nuclear burning, all the way to iron, with gigantic nuclear
detonations seen in the form of quasar outbursts. For a galactic infall rate
of 1M$_{\odot}$/yr, only 3 Myr have to pass for sending the present mass of
Sgr A* into our Galactic center!

\begin{figure}
\centering 
\begin{minipage}[b]{16.cm}
\centering
\includegraphics[width=16.0cm]{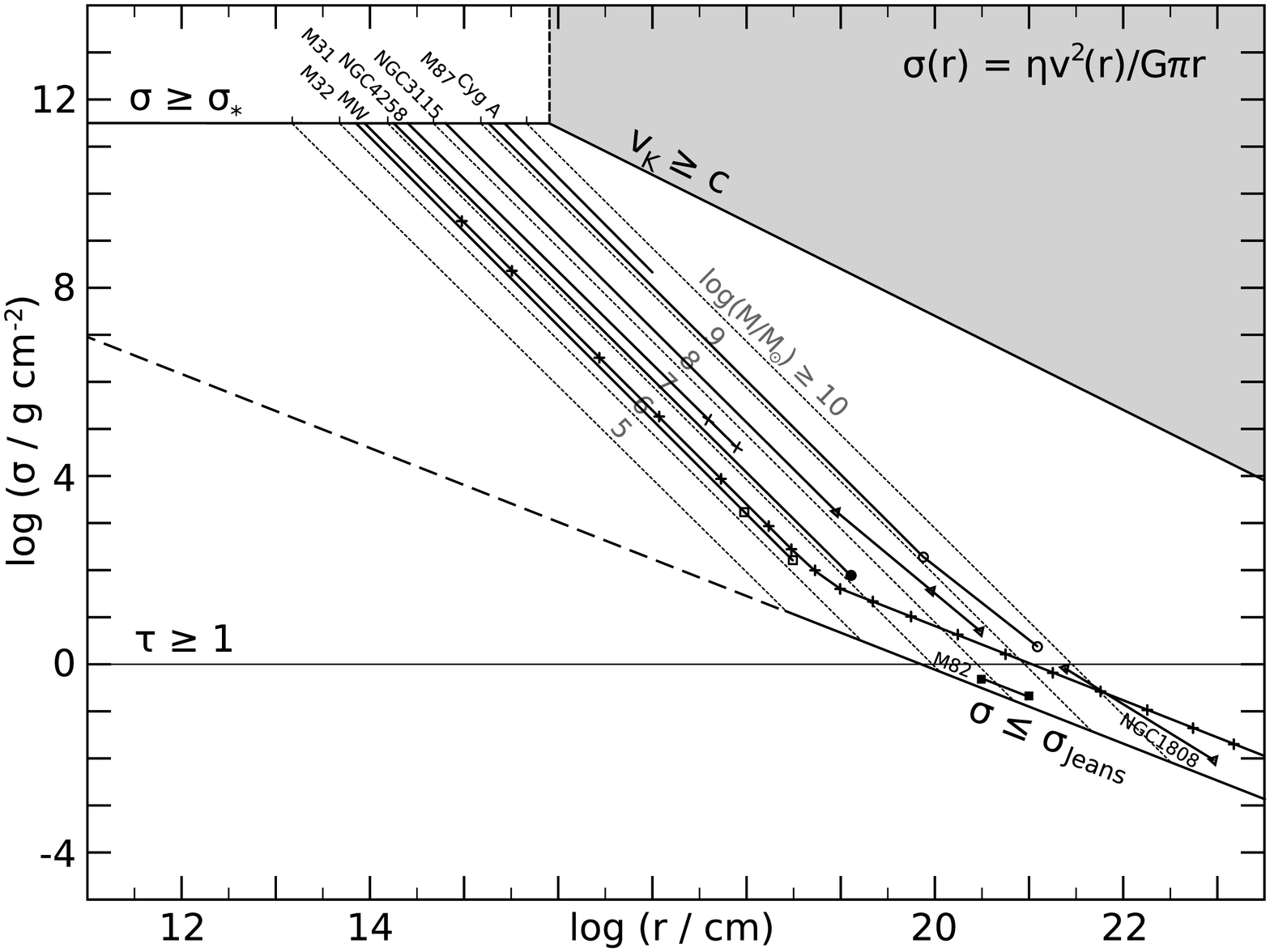} 
\parbox[b]{145mm}{\small \  }
\parbox[b]{145mm}{\small \textbf{FIGURE 5.} Complete rotation curves \ -- \ with $10^{11}$cm $\leq$ $\ r$ $\leq$
$10^{23.5}$cm \ \ -- \ for a representative set of well-sampled galaxies,
taken from (Kundt 2008a). For a better understanding of galactic centers, the
ordinate presents average surface-mass density \ $\sigma(r)$ $\lesssim$
$v^{2}(r)$ $/$ $G$ $\pi$ $r$ \ (instead of rotational velocity $v(r)$) :
Whilst $\sigma(r)$  is tiny in the outer parts of a galaxy, where it is
controlled by Jeans instability (to star formation), it grows considerably
with decreasing $r$, but cannot exceed stellar values ($\sigma_{\ast}%
\approx10^{11.5}$ g/cm$^{2}$), due to pressure forces, hence sets a bound on
revolution speeds near the center. Observations indicate that galaxies have
ringlike domains of insignificant (gravitating) mass density, between
$\gtrsim10^{14}$cm \ and $\lesssim10^{20}$cm, in which their rotation is
solely controlled by the mass of their central engine (CE), and $M(r)$ =
const. Note that the detected CE masses all stay below the BH formation limit
of $10^{10.5}$M$_{\odot}$ -- marked in gray -- beyond which they would enforce
(among others) extremely relativistic galactic revolution speeds. }
\end{minipage}
\end{figure}

Mass-infall rates into the center compensate mass-ejection rates when
integrated over a typical quasar cycle. The hot cores have radial extents
between $10^{16}$cm and $\lesssim10^{14}$cm, vertical extents comparable to
stellar diameters, and evolve chemically during spiral-in of their substratum
(Kundt 2008a). The BDs are somewhat larger in extent than BHs (for the same
mass), and have never reached instability towards gravitational collapse.
During active cycles, their QPO variability timescales show a white power
distribution, with\ an upper break frequency $f$\ of
\begin{equation}
\ f\lesssim3\text{ KHz }(M_{\odot}/M)\label{QPO}%
\end{equation}
found by Remillard and McClintock (2006), which relation holds throughout
more than nine orders of magnitude in mass, from the stellar-mass black-hole
candidates to the most massive (well-sampled) CEs of active galaxies. Famous
examples for (\ref{QPO}) are Sgr A* , with its bursts of duration
$\lesssim20\min$ (for a CE mass of $10^{6.5}$M$_{\odot}$), and RE J1034+396,
with its sampled one-hour quasi periodicity (and mass $10^{7}$M$_{\odot}$),
(Gierli\'{n}ski et al 2008). These preferred (shortest) QPO timescales are
reminiscent of  --  but distinctly longer (10 times) than  --  the innermost
Kepler periods of a BH. To me, they look like magnetospheric oscillation
cycles. (Note that these engines can emit their power above TeV particle
energies! Boosted via magnetic slingshots?).

Why do I mistrust the BH interpretation, (since $\gtrsim$30 years)? As already
explained, I cannot see the holes' formation mode: nature has provided
hurdles, such as centrifugal forces, pressures, and detonations. Fig.5 shows
that the BH rotation curves avoid the (upper right) BH formation regime; they
stay below, in surface-mass density $\sigma$. They would touch it as soon as
galactic revolution speeds, at some inner radius, would reach the speed of
light, (and cause that region to flare!). Moreover, even if a BH had somehow
formed, and grown in mass to some $10^{10}$M$_{\odot}$, how would it interact
with its surroundings? All ambient matter would be sucked into it, true, at
speeds approaching the speed of light. But its tidal forces would be minute,
because its curvature radius has heliospheric size, some $10^{15.5}$cm, too
large to strain, or squeeze the infalling CSM towards significant densities.
That infalling CSM would heat up a bit during its compression, though hardly
above X-ray temperatures, and would moreover (i) reach infinity strongly
redshifted. Such dissipative heating would be (ii) accretion-rate
dependent, scaling as $n^{2}d^{3}x$, hence would tend to zero with a decreasing
mass infall rate. For BHs above $10^{8}M_{\odot}$, (iii) accretion at the
Eddington
rate would require supergalactic mass infall rates, $\dot M > M_{\odot}/yr$.
Earlier estimates (by other people) applied accretion-disk efficiencies,
and considered a 
potential energy of 0.42 $\times$ rest energy at the innermost stable orbit
of a maximally spinning BH. They ignored (iv) an optically thick zone around it,
which would be swallowed whole, and which grows with increasing density $n$. To
me, AGN observations never reveal radiated powers of the CE as large as
$10^{-3}$ of its accreted power, in agreement with above considerations.
Large efficiencies of BH accretion have never been demonstrated.
 
The best-studied CE of all is that of our Milky Way galaxy, Sgr A*, at a
distance of $\lesssim$ 8.0 Kpc, whose spectrum is almost white in power ($\nu
S_{\nu}$ = const) from $10^{12}$Hz up to TeV energies, with an integrated
power of $\gtrsim10^{37}$erg/s which may peak at GeV energies. It shows
simultaneous daily bursts at radio and X-ray frequencies, of duration
$\gtrsim$17min. On 16 Nov. 2007, Frank Eisenhauer told us at Bonn that the
(16yr) Kepler ellipse of star S2 around Sgr A* does not close, by 3$^{0}$,
which indicates the gravitational potential of a massive disk (instead of a
pointlike BH). This indication is supported by a growing mass estimate of Sgr
A* with increasing approach, between 2003 and 2007, from $10^{6.46}$ to
$10^{6.58}$ or even $10^{6.63}$ M$_{\odot}$, depending on the correct distance
to it, which Reinhard Genzel reported as $d$ = 8.33 Kpc (on 9 Jan. 2009). Note
that $d$ is used to convert angular velocities (on the sky) into transverse
velocities in space, whilst it leaves Doppler velocities unaffected; again,
this determination prefers a disklike gravitational potential to the (almost)
Coulomb potential of a BH. These three worries will grow into certainties, or
disappear, with the accumulating number of measurements during the coming years.

An independent signature of the BD character of Sgr A* is its gigantic wind,
seen to blow radial tails from the windzones of $\gtrsim$ 8 nearby stars, at
distances $\lesssim$ lyr, and mapped in the redshifted light of extended
Br$\alpha$, and in the blueshifted light of Br$\gamma$, of mass rate some
$10^{-2.5}M_{\odot}$/yr, and speed $\lesssim10^{3}$Km/s, (Kundt 1990). No hole
can expel more matter than you dump on it.

\section{5. ANOMALOUS\ REDSHIFTS AND\ JETS}

In this last section of `critical thoughts', I dare touching upon one of the
most tenacious worries in Cosmology, shared by Halton Arp, Fred Hoyle,
Geoffrey and Margret Burbidge, Mart\'{i}n L\'{o}pez Corredoira, and a few
others, though ignored by the rest of the community: the many close
associations, in the sky, of objects of vastly differing redshifts, the
phenomenon of the ''anomalous redshifts'' (Hoyle et al 2000, Arp 2008). Are
celestial redshifts always cosmological, or are they occasionally simply
kinematic? In the case of the GRBs, I maintain the latter, since more than 15
years (Kundt 2009). Here, for the first time, I maintain again the latter, for
the ''worst case'' according to Arp, the Seyfert galaxy NGC 7603 and its near
celestial neighbours; cf. L\'{o}pez-Corredoira and Guti\'{e}rrez (2002, 2004).

Fig.6A shows the celestial field around NGC 7603, measuring 2' across. The
bright Seyfert 1 galaxy NGC 7603, of redshift z = 0.029 (and distance
124\ Mpc, for H$_{0}$ = 70 Km/s Mpc), is connected to an object without
emission lines, called NGC 7603B, at (larger) redshift z = 0.057, by a curved
luminous bridge called ''filament'', of absorption redshift z = 0.030. The
filament, in turn, contains two compact emission-line\ `knots' of redshifts z
= 0.243 and 0.391, (from OII, H$\beta$, OIII, OI, NeIII, and H$\alpha$,
corresponding to velocity spreads of $\lesssim$ $10^{3.2}$Km/s), which have
been resolved and mapped in (2004), with FWHM $\gtrsim$0.3''. Conservatively,
the three knotlike objects have been interpreted as HII-galaxies, or NEL
galaxies, at much larger distances.

\begin{figure}
\centering
\begin{minipage}[b]{16.cm}
\centering
\includegraphics[width=15.0cm]{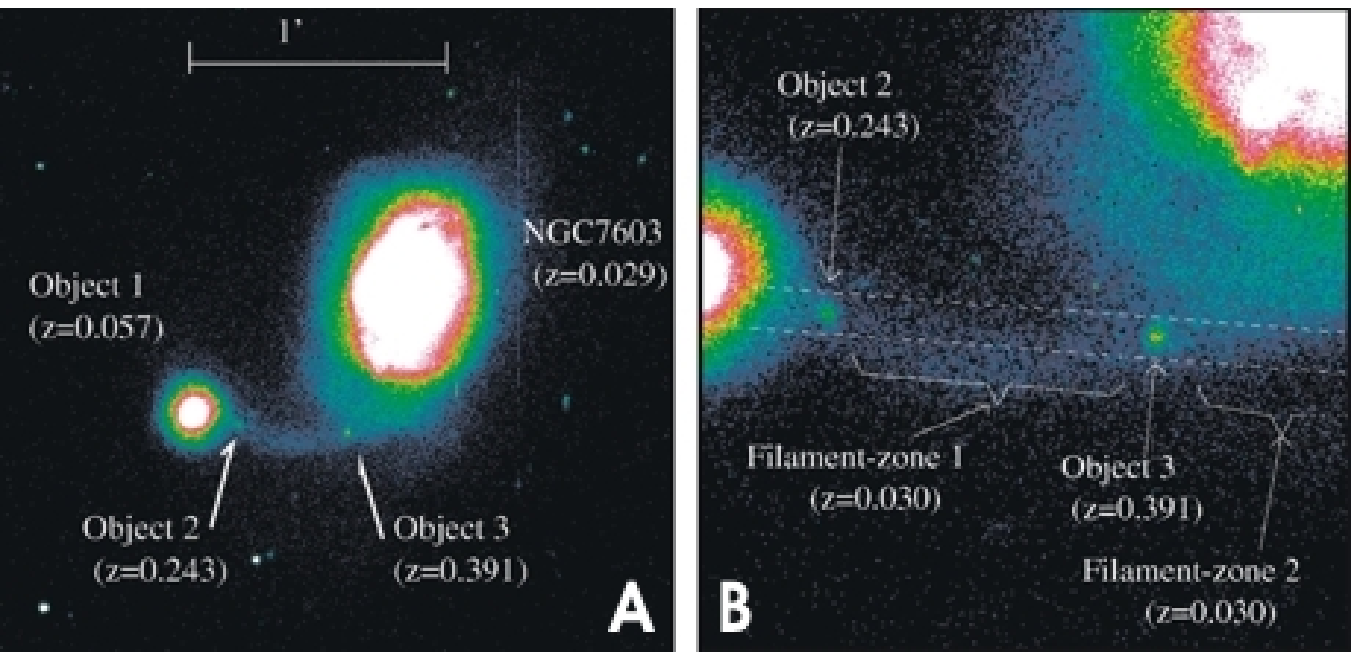}
\parbox{146mm}{\small \  }
\parbox[b]{146mm}{\small \textbf{FIGURE 6.} The celestial neighbourhood of the Seyfert 1 galaxy NGC 7603, taken
from L\'{o}pez-Corredoira and Guti\'{e}rrez (2002). It is connected by a
curved, luminous bridge to an object 1 (also called\ NGC 7603B by them) of
higher redshift but without emission lines. The bridge is slightly redshifted,
again without emission lines, but contains two compact knots -- called objects
2 and 3 in (2002) \ -- \ of considerable redshifts and line-broadening,
corresponding to FWHM speeds of $10^{3.2}$Km/s, which were shown later (2004)
to have FWHM extents of $\gtrsim$0.3''. In the text I interpret NGC 7603B as
the head of a receding jet, whose lobe is seen as the (mildly receding)
bridge, and objects 2 and 3 as (fast) knots swept up by the (pair-plasma) jet. }
\end{minipage}
\end{figure}

But in my 1986 paper with Gopal Krishna, in which we elaborate on the bright
jet source 3C 273, we find an approach velocity c$\beta$ of its head of order
$\beta\approx$ z $\approx$\ 0.7, and a non-detection of its inner part, and of
its expected redshifted lobe. We conclude at an extremely strong, approaching
galactic (pair-plasma) jet propagating through a rather thin circumstellar
medium (CSM), so that its channel-wall material (or head) gets boosted to
transrelativistic speeds. Another such blue-shifted jet source, CGCG 049-033,
has been recently identified by Bagchi et al (2007); its receding lobe is
unseen, most likely for causality reasons.

I therefore like to interpret above NGC 7603 as a radio galaxy of which we see
a receding ''lobe'' -- \ called ''filament'' above \ -- whose red-shifted head
is NGC 7603B. The mildly redshifted lobe contains two fast, more strongly
redshifted knots: the inner one of slightly higher redshift than the outer
one, both of low column density, hence emission-line objects, with a
considerable spread of velocities (caused by the jet's impacting at different
strengths). Both knots and head are formed from ambient (channel-wall)
material swept up by the extremely relativistic pair-plasma jet. Where is the
opposite, blue-shifted lobe (of the twin jet)? It may well be bent around near
the northwestern edge of NGC 7603, with both lobes forming a large ''U ''
(open in\ `downwind' direction). Alternatively, its blueshifted light may already have
passed us. Note that when two objects are fired in opposite directions at
relativistic speeds, a distant observer aligned with them will see the
blueshifted object for a very short time only -- when its flash passes him or
her -- whilst the redshifted object will stay visible for its whole lifetime.
We thus expect to see many more redshifted knots than blueshifted knots,
perhaps 10-times as many; two blueshifted ones were discussed above. I see no
principle difficulty in identifying a number of high-velocity receding
emission-line knots as luminous channel-wall material in receding lobes. Jet
plasma is thought to move at large Lorentz factors ($\gtrsim10^{2}$, Kundt and
Krishna 2004), and occasionally imposes transrelativistic channel-wall speeds.
Redshifts need not always be cosmological.

Why have corresponding blueshifted emission lines never been reported, from approaching jets? 
They may be difficult to detect: The knots and heads of the (relativistic!) jets are expected to emit their synchrotron radiation strongly in forward directions, whereas their (slowly 
moving) channel-wall material should radiate almost isotropically. Consequently, redshifted 
lines should come from a dark sky, whilst blueshifted lines should be superposed on a strong 
synchrotron continuum. Indeed, the radiation received from the blueshifted hotspots in 3C 33, 
Pictor A, and others may well be such superpositions: Simkin (1986), Simkin et al (1999), 
Tingay et al (2008). An absence of reports need not mean an absence of detections.                                                                                                                                                              
\begin{theacknowledgments}
My warm thanks go to G\"{u}nter Lay and Ingo Thies for help with the electronic data
handling, and to Mario Novello for having invited me to his open-minded School.
Heinz Andernach helped me when searching for the missing blueshift, and Gernot Thuma and Hans Baumann improved the manuscript.
\end{theacknowledgments}


\newpage

\begin{thebibliography}{99}
\bibitem{Arp (2008)}Arp, H.C., Scientific and political elites in western
democracies, in: \textit{Against the Tide}, eds. M.L. Corredoira and C.C.
Perelman, Universal Publishers, 117-128, 2008.

\bibitem{Bagchi et al (2007)}Bagchi, J., Gopal-Krishna, Krause, M., and Joshi,
S., A giant radio jet ejected by an ultramassive black hole in a single-lobed
radio galaxy, \textit{Astrophys. J.} 670, L85-L88, 2007.

\bibitem{Belinski (2006)}Belinski, V.A., On the existence of black hole
evaporation yet again, \textit{Physics Letters A} 354, 249-257, 2006.

\bibitem{Boon and Tsallis (2005)}Boon, J.P., and Tsallis, C., Nonextensive
statistical mechanics, \textit{europhysicsnews} 36/6, 185-186, 2005.

\bibitem{Carroll (2008)}Carroll, S.M., The Cosmic Origins of Time's Arrow,
\textit{Scientific American}, June, 26-34, 2008.

\bibitem{Crawford (2008)}Crawford, D.F., \textit{Curvature Cosmology}, Brown-Walker Press, 2008.

\bibitem{Ellis 2008)}Ellis, G., Patchy solutions, \textit{Nature} 452,
158-160, 2008.

\bibitem{Fixsen (2003)}Fixsen, D.J., The spectrum of the cosmic microwave
background anisotropy from the combined COBE, FIRAS, and WMAP observations,
Astrophys. J. 594, L67-70, 2003.

\bibitem{Gierlinski et al (2008)}Gierli\'{n}ski, M., Middleton, M., Ward, M.,
and Done, Ch., A periodicity of $\sim$1 hour in X-ray emission from the active
galaxy RE J1034+396, \textit{Nature} 455, 369-371, 2008.

\bibitem{Hawking (1974)}Hawking, S.W., Black hole explosions? \textit{Nature}
248, 30-31, 1974.

\bibitem{Hawking (1975)}Hawking, S.W., Particle creation by black holes,
\textit{Commun. Math. Phys.} 43, 199-220, 1975.

\bibitem{Hoyle et al (2000)}Hoyle, F., Burbidge, G., and Narlikar, J.V.,
\textit{A Different Approach to Cosmology}, Cambridge Univ. Press, 2000.

\bibitem{Kanekar et al (2005)}Kanekar, N., et al (10 authors), Constraints on
Changes in Fundamental Constants from a Cosmologically Distant OH Absorber or
Emitter, \textit{Phys. Rev. Lett.} 95, 261301(4), 2005.

\bibitem{Kundt (1972)}Kundt, W., Global Theory of Spacetime, in:
\textit{Proceedings of the 13th Biennial Seminar of the Canadian Mathematical
Congress (at Halifax, in August 1971)}, ed R. Vanstone, Montreal, Vol. 1,
93-133, 1972.

\bibitem{Kundt (1976)}Kundt, W., Entropy production by black holes,
\textit{Nature} 259, 30-31, 1976.

\bibitem{Kundt (1990)}Kundt, W., The Galactic Centre, \textit{Astrophys. and
Space Science} 172, 109-134, 1990.

\bibitem{Kundt (2002)}Kundt, W., Radio galaxies powered by burning disks,
\textit{New Astronomy Reviews} 46, 257-261, 2002.

\bibitem{Kundt (2005)}Kundt, W., \textit{Astrophysics, A New Approach},
Springer, 2005.

\bibitem{Kundt (2007)}Kundt, W., Fundamental Physics, \textit{Foundations of
Physics} 37, No. 9, 1317-1369, 2007.

\bibitem{Kundt (2008a)}Kundt, W., The Proposed Black Holes around us, in:
\textit{The 11th Marcel Grossmann Meeting on General Relativity and
Gravitation}, World Scientific, 1529-1536, 2008a.

\bibitem{Kundt (2008b)}Kundt, W., Supernovae, their functioning, lightcurves,
and remnants, \textit{New Astronomy Reviews} 52, 364-369, 2008b.

\bibitem{Kundt (2009)}Kundt, W., The sources of the Cosmic Rays, and of the
Gamma-Ray Bursts, after more than \{40,30\}years of deliberation,
\textit{Chinese J. of Astronomy and Astrophysics}, in print, 2009.

\bibitem{Kundt and Krishna (1986)}Kundt, W., and Krishna, G., The Jet of the
Quasar 3C 273, \textit{J. Astrophys. Astr.} 7, 225-236 , 1986.

\bibitem{Kundt and Krishna (2004)}Kundt, W., and Krishna, G., The Physics of E
x B-Drifting Jets, \textit{J. Astrophys. Astr.} 25, 115-127, 2004.

\bibitem{Layzer (1990)}Layzer, D., \textit{Cosmogenesis}, Oxford Univ. Press, 1990.

\bibitem{Leblanc (2002)}Leblanc, Y., The Quantum Black Hole Problem,
\textit{eFieldTheory.COM/articles}/021201, 2002.

\bibitem{Lopez-Corredoira (2002)}L\'{o}pez-Corredoira, M., and Guti\'{e}rrez,
C.M., Two emission-line objects with z $>$ 0.2 in the optical filament apparently connecting the Seyfert galaxy NGC 7603
to its companion, \textit{A \& A} 390, L15-L18, 2002.

\bibitem{Lopez-Corredoira (2004)}L\'{o}pez-Corredoira, M., and Guti\'{e}rrez,
C.M., The field surrounding NGC 7603: Cosmological or non-cosmological
redshifts? \textit{A \& A} 421, 407-423, 2004.

\bibitem{Marconi and Hunt (2003)}Marconi, A., and Hunt, L.K., The relation
between black hole mass, bulge mass, and near-infrared luminosity,
\textit{Astrophys. J.} 589, L21-L24, 2003.

\bibitem{McGaugh and de Blok (1998)}McGaugh, S.S., and de Blok, W.J.G.,
Testing the dark matter hypothesis with low surface brightness galaxies and
other evidence, \textit{Astrophys. J.} 499, 41-65, 1998.

\bibitem{Remillard and McClintock (2006)}Remillard, R.A., and McClintock,
J.E., X-ray properties of black-hole binaries, \textit{Ann. Rev. A \& A} 44,
49-92, 2006.
\bibitem{Simkin (1986)}Simkin, S.M., Optical Spectroscopy of the Southwest
Radio Lobe in 3C 33, \textit{Astrophys. J.} 309, 100-109, 1986.

\bibitem{Simkin et al (1999)}Simkin, S.M., Sadler, E.M., Sault, R., Tingay,
S.J., and Callcut, J., Pictor A (PKS 0518-45): From Nucleus to Lobes, \textit
{Astrophys. J. Suppl.} 123, 447-465, 1999. 

\bibitem{Sorrell (2006)}Sorrell, W.H., The cosmic age crisis, the Hubble
constant, and the cosmic microwave background radiation in a non-expanding
universe, \textit{two preprints}, St. Louis, 2006.

\bibitem{Tingay et al (2008)}Tingay, S.J., Lenc, E., Brunetti, G., and Bondi,
M., A high resolution view of the jet termination shock in a hot spot of the
nearby radio galaxy Pictor A, implications for X-ray models of radio galaxy
hot spots, \textit{Astrophys. J.} 136, 2473-2482, 2008.

\bibitem{Vestergaard et al (2008)}Vestergaard, M., Fan, X., Tremonti, C.A.,
Osmer, P.S., and Richards, G.T., Mass functions of the active black holes in
distant quasars from the Sloan Digital Sky Survey Data Release 3,
\textit{Astrophys. J.} 674, L1-L4, 2008.

\bibitem{Weekes (2007)}Weekes, T., Photons from a hotter hell, \textit{Nature}
448, 760-762, 2007.

\bibitem{Wiltshire (2007a)}Wiltshire, D.L., Dark Energy without Dark Energy,
\textit{arXiv}:07123984, 2007a.

\bibitem{Wiltshire (2007b)}Wiltshire, D.L., Cosmic clocks, cosmic variance,
and cosmic averages, \textit{New J. Physics} 9, 377-449 (2007b), or: arXiv: gr-qc/072082v4.
\end{thebibliography}
\end{document}